\pdfoutput=1

\documentclass[11pt]{article}

\usepackage[final]{acl}

\usepackage{times}
\usepackage{latexsym}
\usepackage{amsmath}

\usepackage[T1]{fontenc}

\usepackage[utf8]{inputenc}

\usepackage{microtype}

\usepackage{inconsolata}

\usepackage{graphicx}

\title{Semantic Search for Information Retrieval}

\author{
  Kayla Farivar \\
  University of California, Berkeley \\
  \texttt{kaylafari@berkeley.edu}
}

\begin{document}
\maketitle
\begin{abstract}
Information retrieval systems have progressed notably from lexical techniques such as BM25 and TF-IDF to modern semantic retrievers. This survey provides a brief overview of the BM25 baseline, then discusses the architecture of modern state-of-the-art semantic retrievers. Advancing from BERT, we introduce dense bi-encoders (DPR), late-interaction models (ColBERT), and neural sparse retrieval (SPLADE). Finally, we examine MonoT5, a cross-encoder model. We conclude with common evaluation tactics, pressing challenges, and propositions for future directions.
\end{abstract}

\section{Introduction to Information Retrieval}
Information retrieval (IR) refers to returning relevant information based on a human-made query and a knowledge base to search (generally a large collection of documents). After taking a source of knowledge and a pre-made query, an IR system uses a retriever to return relevant information from the knowledge base. A retriever picks which information to return based on the semantic similarity of the query and chunks of each document \cite{muller-gurevych-2009-study}. Semantic similarity measures the closeness of the contextual meaning of two pieces of text. This is important due to the volatility of language - a small change in a sentence can completely shift the meaning of the sentence. IR is used for various applications. A search engine may use a reranking layer to give a user multiple options of what documents to look at, while another system may use the information to generate an answer from an LLM \cite{jiang-etal-2023-active}.

\section{History}
BM25 (Best Match 25) \cite{conf/trec/RobertsonWJHG94} is a purely lexical scoring function popularly used for ranking in information retrieval systems. Although this function used no semantic properties, it’s still highly regarded and used as a baseline against modern-day retrievers. BM25 is highly inspired by another formula for ranking called TF-IDF (Term Frequency - Inverse Document Frequency). Both translate documents into a Bag-of-words vector. Bag-of-words represents documents as a vector whose dimension size is of an all-encompassing set of vocabulary, each entry holding the word's frequency in a specific document (or zero if the term is absent) \citep{salton_termweighting_1988}. The returning vector explains the meaning of a document through the terms that could have been mentioned and the terms that had been mentioned. BM-25 deviates by introducing diminishing returns for term frequency, so as a term appears more and more often, it doesn’t increase how much it matters. Additionally, it standardizes documents so they aren’t awarded for being longer. 

In the early 2010s, semantic embeddings were popularized with the introduction of word2vec \citep{mikolov2013efficientestimationwordrepresentations}. Word2vec uses local context to determine similarity. Assume if we often saw sentences about lions, we could replace the word lion with the word giraffe. Word2vec would then correctly assume lions and giraffes have a close relationship as animals. BM25 kept steady as the strongest scoring function through the innovations around embeddings, the most famous being BERT \citep{devlin-etal-2019-bert}, until ORQA \cite{lee-etal-2019-latent} made a dense retriever that was very competitive with BM25’s scores. ORQA’s designs utilized BERT, a bidirectional encoder that provides highly contextual embedding outputs, and provided high motivation for semantic search \cite{yates-etal-2021-pretrained}.

\section{Dense Bi-Encoder Models}
\citet{karpukhin-etal-2020-dense} introduced one of the first Dense Bi-Encoders inspired by ORQA’s designs. The model uses BERT as an encoder and finds the top-k most similar queries to document pairs with a near-real-time algorithm from FAISS \citep{douze2025faisslibrary}, which uses complex index structures to speed up algorithms. With BERT, the last token in a set of embeddings is the [CLS] token, which includes an understanding of every token before it. They chose their formula for similarity as the dot product between the [CLS] tokens for the query and document. The bulk of their work has to do with their chosen method of negative training. Within training, there is one correct passage for a given query, negative training refers to the samples given to the model that are not correct. They tested a few collections of negative samples. First, they tried a series of three collections: random incorrect passages from their data, then only passages that BM25 retrieved but weren't correct, and lastly passages that were only correct for other sample questions. These sets of negative samples comparatively had little to no impact on their top-k accuracy. Next, they tried only using passages that were correct for other sample questions in the same batch during training and found a large improvement in their results. Further trialing found that adding one of BM25’s hard negatives (passages BM25 rated highly but were incorrect) to their samples improved their accuracy further (and adding more than one sample made little to no difference). The resulting model generalized well (small decrease in accuracy when trying independent sources), and its improved design significantly beat BM25 with a top-5 accuracy of 65.2

\section{Late-Interaction Models}
ColBERT \citep{khattab2020colbertefficienteffectivepassage} is a late-interaction model that gets an embedding per token for the input text. ColBERT adds an interesting feature in training called query augmentation. This adds padding to queries with the special [MASK] token. In BERT, this token is usually used for masking words in training texts, and by predicting the masked words, it helps BERT learn what words mean through context. With ColBERT, this additional padding allows the model to imagine what words might exist in the query around it. For example, if a query included the terms “American Independence Day,” it might imagine similar terms like “Fourth of July” or “U.S national holiday” to add near it. This allows for the query to be stronger and find more passages similar to it. For the separate document encoder, there is no padding, but as usual, the text is tokenized and an embedding is made for each resulting token. With two sets of query and document embeddings, a score is computed by finding the maximum value dot product of each query token’s embedding with any embedding from the document’s tokens. In ColBERT \cite{santhanam-etal-2022-colbertv2}, they speed up this process and add token weighting to the query to show stronger importance to select tokens. ColBERT's scoring allows for a high precision because one vector doesn't collect as many nuances and details, although it can be memory-intensive due to the count of tokens in queries and documents. It’s recommended in practice if the amount of documents that need processing is high, ColBERT should be used as a top-k reranking system. The idea is to use a primary retriever like BM25 to get the top-k documents and then feed that smaller k documents into ColBERT.

\section{Neural Sparse Retrieval}
SPLADE \cite{formal2021spladesparselexicalexpansion} is a neural sparse retriever that wanted to take more inspiration from BM25 and move away from dense embeddings in DPR retrievers. BM25 used a very sparse bag-of-words vector. SPLADE wanted to bridge the interpretability of sparse embeddings while training with neural networks like DPR models. To achieve this, SPLADE gets embeddings with BERT (or more commonly at this age, DistilBERT \citep{sanh2020distilbertdistilledversionbert} due to its balance of efficiency and performance) for each token in the query, and it then reuses BERT's masked language modeling head to get logits over the entire vocabulary per token. Logits represent how likely each term is to appear there, and they give an idea of the importance of each term. SPLADE then uses a log saturation function to add up each of the logits into one score. The purpose of using log saturation $ \left( \sum (\log (1+w) \text{ where w = logit weight} \right) $  instead of a simple sum is to suppress large values. This process results in a vocabulary-length vector of various logits indicating its highest predicted relevance. ReLU is then used to remove all negatives and create the final sparse vector. The FLOPS regularizer was also introduced to penalize the loss for frequently active terms across documents. The same actions are repeated for the document with the same encoder, and the dot product of the two is used as their similarity score. 

A year after SPLADE's introduction, a few extra optimizations were introduced for SPLADE v2 \citep{formal2021spladev2sparselexical}. SPLADE v2 further enforces sparsity by updating its FLOPS regularizer to penalize the loss for having too many non-zero outputs. The rest of their optimizations were made into optional settings with a differently named model for each. Inspired by other models (including ColBERT), SPLADE-max uses max pooling to aggregate the logits rather than the previous log saturation sum. SPLADE-doc is a new encoder specifically for documents. When the query is encoded, the system may mark tokens that weren’t exactly in the query as relevant tokens and give each relevant token a weight. This allows for a more versatile search query. The new SPLADE-doc is similar to before, but only tokens that were in the document can have a weight, and all of the weights are set to be equal. This optimization was purely made for efficiency and had little effect on the accuracy. Lastly, they took a new approach called model distillation \citep{hofstätter2021improvingefficientneuralranking}. The idea is to take a highly accurate model (in this case, a cross-encoder) and train triplets (a query, a correct document, and an incorrect document), get this model to score new triplets, and train a lighter weight model (SPLADE in this case) to output these same scores. This allows for a teacher (a cross encoder) to “teach” the student (SPLADE). This new model is called DistilSPLADE-max.

\section{Cross-Encoder Models}
Information Retrieval was dominated by BERT pretrained models until 2020 \citep{wang2024utilizingbertinformationretrieval}, when \citet{nogueira-etal-2020-document} made the cross-encoder MonoT5. T5 \citep{raffel2023exploringlimitstransferlearning} is an encoder-decoder transformer, the encoder takes in a text input and the decoder is run autoregressively to produce a text output. Simplified, T5 pretrains very similarly to BERT's masking training, but with multiple tokens being masked. T5 uses special tokens like <x> <y> <z>, for example, and masks tokens; the model's job is then to predict what goes under each token. This helps the model learn the meaning of words in the context of a sentence, but also learn how to generate context. MonoT5 took in the sequence “Query: q Document: d Relevant: “ (where q and d were the respective documents) and outputed the sequence “true” or “false”. MonoT5 reframed information retrieval and helped popularize using “target words” (like true or false in this instance) to transform sequence-to-sequence models into acting similarly to most other models. To simply get the probabilities of each document, they apply softmax to the logits of the tokens “true” and “false”. This works very well because the target words “true” and “false” are only one token, which allows them to extract the probability of the document. The steps to finetune T5 into MonoT5, can be repeated with larger versions of t5 \citep{ni-etal-2022-large}, GPT 3.5, GPT 4.0, or other LLMS.

\section{Benchmarking and Evaluation}
The most popular evaluation dataset is MS MARCO (MiscroSoft MAchine Reading COmprehension) \citep{bajaj2018msmarcohumangenerated}, debuting in almost every modern paper on information retrieval. The expansive dataset scraped Bing to produce over a million anonymized questions, each with a human-generated answer, and almost 9 million passages. Their team also created a leaderboard with bootstrapping and private queries to validate performance. Although the MS MARCO leader board helps establish a history of improvement, other benchmarks like BEIR (Benchmarking-IR) \citep{thakur2021beirheterogenousbenchmarkzeroshot} should be used in practice to avoid leaderboard chasing and prove well-roundedness \citep{craswell2021msmarcobenchmarkingranking}.

\section{Conclusion and Future Directions}
In summary, semantic search retrievers overtook the lexical innovations of BM25 at their place in information retrieval. These innovations have allowed for the creation of real-time RAG-based \citep{lewis2021retrievalaugmentedgenerationknowledgeintensivenlp} virtual assistants, enhanced search engines, and stronger data analysis for personalized recommendations and real-time information access. While great progress has been made, there are significant challenges that persist. English-centric datasets overlook multilingual compatibility and low-resource languages \citep{yang-etal-2024-language-bias}. Further research is necessary to allow for the collective benefits of technological advancement.

\bibliography{custom, anthology}

\end{document}